\documentclass[twocolumn]{aastex631}
\usepackage{graphicx}
\usepackage{txfonts}
\usepackage{silence}
\usepackage{lastpage}
\usepackage{array}
\usepackage{makecell}
\usepackage{tabularx}

\begin{document}

   \title{The connection between galaxy mergers, star formation and AGN activity in the HSC-SSP}

   \author[0000-0002-8432-6870]{Kiyoaki Christopher Omori
          }
  \affiliation{Department of Astronomy and Physics and Institute for Computational Astrophysics, Saint Mary's University
  923 Robie St
  Halifax, NS B3H 3C3, Canada\\
  }
  \affiliation{Research Center for Space and Cosmic Evolution, Ehime University, 2-5 Bunkyo-cho, Matsuyama, Ehime 790-8577, Japan\\}
  \author{Connor Bottrell}
  \affiliation{International Centre for Radio Astronomy Research, University of Western Australia
  35 Stirling Hwy
  Crawley, WA6009, Australia\\}
    \author{Sabine Bellstedt}
  \affiliation{International Centre for Radio Astronomy Research, University of Western Australia
  35 Stirling Hwy
  Crawley, WA6009, Australia\\}
  \author{Aaron S. G. Robotham}
\affiliation{International Centre for Radio Astronomy Research, University of Western Australia
  35 Stirling Hwy
  Crawley, WA6009, Australia\\}
  \author{Hassen M. Yesuf}
  \affiliation{Key Laboratory for Research in Galaxies and Cosmology, Shanghai Astronomical Observatory, Chinese Academy of Sciences, 80 Nandan Road, Shanghai 200030, China\\}
\affiliation{Kavli Institute for the Physics and Mathematics of the Universe (WPI), The University of Tokyo Institutes for Advanced Study (UTIAS), The University of Tokyo, 5-1-5 Kashiwanoha, Kashiwa-shi, Chiba 277-8583, Japan\\}
\author{Andy D. Goulding}
\affiliation{Department of Astrophysical Sciences, Princeton University, 4 Ivy Lane, Princeton, NJ 08544, USA\\}
\author{Marcin Sawicki}
  \affiliation{Department of Astronomy and Physics and Institute for Computational Astrophysics, Saint Mary's University
  923 Robie St
  Halifax, NS B3H 3C3, Canada\\
  }
\author{Tohru Nagao}
\affiliation{Research Center for Space and Cosmic Evolution, Ehime University, 2-5 Bunkyo-cho, Matsuyama, Ehime 790-8577, Japan\\}
\author{Tsutomu T. Takeuchi}
\affiliation{Division of Particle and Astrophysical Science, Nagoya University, Furo-cho, Chikusa-ku, Nagoya 464--8602, Japan\\}
\affiliation{The Research Centre for Statistical Machine Learning, the Institute of Statistical Mathematics, 10--3 Midori-cho, Tachikawa, Tokyo 190--8562, Japan\\}

 
\begin{abstract}
   {Internal gas inflows driven by galaxy mergers are considered to enhance star formation rates (SFR), fuel supermassive black hole growth and stimulate active galactic nuclei (AGN). However, quantifying these phenomena remains a challenge, due to difficulties both in classifying mergers and in quantifying galaxy and AGN properties. We quantitatively examine the merger-SFR-AGN connection using Hyper Suprime-Cam Subaru Strategic Program (HSC-SSP) galaxies using novel methods for both galaxy classification and property measurements.}
   {Mergers in HSC-SSP observational images are identified through fine-tuning Zoobot, a pretrained deep representation learning model, using images and labels based on the Galaxy Cruise project. We use galaxy and AGN properties that were produced by fitting Galaxy and Mass Assembly (GAMA) spectra using the SED fitting code ProSpect, which fits panchromatically across the far-UV through far-infrared wavelengths and obtains galaxy and AGN properties simultaneously.}
{Little differences are seen in SFR and AGN activity between mergers and controls, with $\Delta \mathrm{SFR}=-0.009\pm 0.003$ dex, $\Delta f_{\mathrm{AGN}}=-0.010\pm0.033$ dex and $\Delta L_{\mathrm{AGN}}=0.002\pm0.025$ dex. After further visual purification of the merger sample, we find $\Delta \mathrm{SFR}=-0.033\pm0.014$ dex, $\Delta f_{\mathrm{AGN}}=-0.024\pm0.170$ dex, and $\Delta L_{\mathrm{AGN}}=0.019\pm0.129$ dex for pairs, and $\Delta \mathrm{SFR}=-0.057\pm0.024$ dex, $\Delta f_{\mathrm{AGN}}=0.286\pm0.270$ dex, and $\Delta L_{\mathrm{AGN}}=0.329\pm0.195$ dex for postmergers. These numbers suggest secular processes being an important driver for SF and AGN activity, and present a cautionary tale when using longer timescale tracers.}

\end{abstract}
\keywords{Active galaxies --
                Interacting galaxies
               }

%

\section{Introduction}
\label{sec:intro}

   Almost all massive galaxies host a central supermassive black hole (SMBH, \citealt{1995ARA&A..33..581K,2013ARA&A..51..511K}). SMBHs can grow through merger and accretion events. When gas joins the accretion disk surrounding a SMBH, a large amount of energy is released as radiation. Galaxies with this bright central emission are called active galactic nuclei (AGN) hosts. 

   One of the avenues that is considered to be associated with such inflow and the subsequent ignition of AGNs is the process of galaxy interactions and mergers \citep{2005Natur.433..604D, 2008ApJS..175..356H,2023MNRAS.519.4966B,2023MNRAS.519.2119Q,2024MNRAS.528.5864B,2024MNRAS.533.3068B}. When  galaxies interact and merge, strong gravitational torques will result in dissipation of angular momentum, leading to gas infall into the central regions of galaxies. Such infall not only feeds SMBH growth and fuels subsequent AGN activity, but can trigger strong star formation activity and starbursts \citep{2000ApJ...530..660B,2005ASSL..329..143S,2009ApJ...694L.123K,2009PASJ...61..481S,2014MNRAS.442L..33R,2015MNRAS.446.2038R,2019MNRAS.490.2139R}. The consumption of gas by star formation, in addition to heating, outflows, and feedback caused by star formation and AGN activity, will lead to eventual quenching of the galaxy \citep{2022MNRAS.516.4354W,2023MNRAS.519.2119Q,2024arXiv241006357E}.
   While it has been shown that merger activity triggers star formation and is an efficient pathway for SMBH growth, quantitative conclusions on the relative role of mergers on star formation and AGN activity are still contested. For star formation, while many studies have found that star formation is enhanced among mergers and other morphologically disturbed galaxies \citep{2008ChJAS...8...77B, 2008AJ....135.1877E,2011MNRAS.412..591P,2013MNRAS.433L..59P,2013MNRAS.430.1901H,2015MNRAS.448.1107M,2016MNRAS.462.2418S,2019MNRAS.482L..55T, 2021ApJ...923..205Y,2024MNRAS.527.6506B}, other studies do not find a pronounced increase in star formation rate among mergers \citep{2019A&A...631A..51P, 2023ApJ...953...91L}. With regard to AGNs, observational studies show that close galaxy pairs and galaxies with disturbed or asymmetric morphologies are (a) more likely to host AGN and (b) have higher AGN luminosities than non-merger control galaxies \citep{2011MNRAS.418.2043E, 2011ApJ...743....2S, 2014AJ....148..137L,2014MNRAS.441.1297S,2017MNRAS.464.3882W, 2018PASJ...70S..37G,2023MNRAS.521.5272T}. Similarly, a number of studies have found an increased AGN fraction among merging systems compared to non-merging systems \citep{2012ApJ...758L..39T, 2012A&A...540A.109S,2015ApJ...814..104K,2019MNRAS.487.2491E}. However, a number of studies find no clear connection between merger and AGN activity \citep{2005ApJ...627L..97G,2009ApJ...691..705G,2011ApJ...726...57C,2012ApJ...744..148K,2014MNRAS.439.3342V, 2016ApJ...830..156M, 2020A&A...637A..94G,2021ApJ...909..124S}. Other works have found a merger-AGN connection, but only among the most luminous of AGNs \citep{2008ApJ...674...80U,2012ApJ...758L..39T,2014A&A...569A..37M,2015ApJ...806..218G,2015ApJ...804...34H,2018MNRAS.476.2308W,2020MNRAS.494.5713M}, while others do not agree with even this conclusion, not finding any merger-AGN connection even among the brightest AGNs \citep{2017MNRAS.470..755H,2017MNRAS.466..812V}.

   One possible reason for the variety of conflicting results is the identification method, for both mergers and AGN hosts. Mergers can be identified through a variety of methods. Some studies identify mergers based on spectroscopic close pair matching, where galaxies with similar spectroscopic redshifts and close distances are considered mergers \citep{2011MNRAS.418.2043E,2011ApJ...743....2S}. Others classify galaxies based on galaxy morphologies, such as non-parametric summary statistics, for example the CAS (concentration, asymmetry, smoothness) parameters \citep{2003ApJS..147....1C}, and Gini-$M_{20}$ \citep{2004AJ....128..163L}. A combination of these statistics can be used to create a \textit{n}-dimensional feature space for merger classification \citep{2018PASJ...70S..37G}. Other studies employ pure visual classification \citep{2008ApJ...674...80U,2011ApJ...726...57C,2014MNRAS.441.1297S,2024PASJ...76..950I}, where merger classification is made by specialists or citizen scientists on observational images. The merger identification method can greatly alter the merger sample, even on similar datasets. This is because different methods can be sensitive to different types of mergers. Spectroscopic pair matching can identify early stage mergers with two nuclei, but fails to recognize post-coalescence or late stage mergers with only one nucleus. Summary statistics such as the CAS parameters can identify many early stage and post-coalescence mergers due to their disturbed morphologies, but may fail in other signatures and are highly sensitive to galaxy physical properties such as gas content \citep{2008MNRAS.391.1137L}. More recently, with advancements in machine learning and deep learning techniques, the use of convolutional neural networks (CNNs) has become increasingly frequent for morphological classifications. Not only are CNN-based representation models able to extract far more morphological information compared to single-parameter or summary statistics \citep{2024MNRAS.528.7411E}, but they are also time and human-resource efficient compared to human powered methods, especially in the current era of large-scale astronomical datasets. CNNs have been used for merger galaxy classification in previous studies \citep{2021MNRAS.504..372B,2022MNRAS.514.3294B,2023A&A...679A.142O} at varying levels of accuracy. 

   With regard to AGN identification, there also exists difficulties. Factors such as dust and the orientation of the accretion disk can obscure galaxy light, making them optically faint \citep{2005ApJ...628..604Y, 2005ApJ...622L.105H,2008ApJ...677..943D,2017ApJ...835...36T}. Dust absorbs and is heated by ultraviolet and optical emission, and the subsequent thermal heating is emitted at infrared wavelengths \citep{1996ARA&A..34..749S, 2015ApJ...814....9K}. Further, emission from AGN at rest-frame UV-optical wavelengths can be difficult to distinguish from the emission of the galaxy itself \citep{2017A&A...604A..99C, 2022MNRAS.509.4940T}. AGN emission is also visible at X-ray \citep{1991ApJ...380L..51H} and radio \citep{1984RvMP...56..255B} wavelengths. As a result, a panchromatic assessment of AGNs is required to characterize AGN activity, and there exists a wide variety of identification methods, such as emission line diagnostics (BPT diagrams, \citealt{1981PASP...93....5B}), colour-selection \citep{2012ApJ...748..142D}, X-Ray selection \citep{2016ApJ...817...34M}, or radio selection \citep{2016ApJ...817...34M}. Each of these different methods can identify AGNs at varying stages of its lifetime \citep{1988ApJ...325...74S}, or with different physical properties, which can lead to widely varying selections \citep{2013ApJ...764..176J,2015ApJ...811...26T}. In addition, the difference in timescale between AGN and merger activities can lead to mixed results. The timescale of AGN activity can range from days to $10^{7-8}$ years, depending on the type of AGN and the timescale over which activity is defined \citep{1996ApJ...470..322C,1996ASPC..110....3S,2004MNRAS.351..169M,2015MNRAS.451.2517S,2021Sci...373..789B}.These timescales are typically far shorter compared to the observability timescale of merger signatures, which can last from 0.2 to several Gyr \citep{2008MNRAS.391.1137L,2010MNRAS.404..575L,2010MNRAS.404..590L}. As such, even if every merger triggered AGN activity, it is not expected to see a 100\% AGN fraction in observational merger samples as the timescales in which SMBHs flip between active and inactive are much shorter than the timescale of the merger process.

   In this work, we identify galaxies from the multi-tiered, wide-field, multi-band imaging survey Hyper Suprime-Cam Subaru Strategic Program (HSC-SSP; \citealt{2018PASJ...70S...8A}). Mergers are identified from HSC-SSP \textit{gri} optical images using the deep representation learning model, Zoobot \citep{Walmsley2023}. The pre-trained Zoobot model is fine-tuned to classify mergers using labels from the Galaxy Cruise \footnote{\url{https://galaxycruise.mtk.nao.ac.jp/en/}} citizen science project \citep{2023PASJ...75..986T}. The fine-tuned model is able to diversely identify merger galaxies at varying merger stages and mass ratios. We use galaxy stellar masses, star formation rates, and AGN properties obtained by the SED fitting code \textsc{ProSpect} \citep{2020MNRAS.495..905R, 2021MNRAS.505..540T,2022MNRAS.509.4940T}, which simultaneously recovers galaxy and AGN properties by panchromatically fitting broadband light from the far-ultraviolet (FUV) to far-infrared (FIR). The simultaneous extraction of galaxy and AGN properties is able to avoid systematics that arise when separately measuring them, such as overestimation of stellar masses and SFRs \citep{2023ApJ...959L..18D}. This is the first work where the merger-AGN connection is investigated with galaxy and AGN properties simultaneously obtained using a self-consistent approach. We combine the two methods that identify diverse mergers and AGN to investigate the relation between merger, star formation, and AGN activity.

   This paper is structured as follows. Section \ref{Section:Samples} highlights our sample and methods for merger selection and obtaining galaxy properties. Section \ref{section:Results} shows our results, and Section \ref{section:Discussion} discusses them. We conclude our work in Section \ref{section:Conclusions}.

     For this paper, we adopt a $\Lambda$CDM model with the following cosmological parameters: $H_0 = 70\textrm{ km } \textrm{s }^{-1} \textrm{ Mpc}^{-1}$, $\Omega_{\Lambda}=0.7$, $\Omega_{M}=0.3$.

\section{Sample Selection and Methods}
\label{Section:Samples}

In this work, we use galaxies from the multi-tiered, wide-field, multi-band imaging survey HSC-SSP, which is observed using the Subaru 8.2 m telescope on Manuakea in Hawaii. Further details about the HSC-SSP survey, the instrumentation used, and its various techniques are available in \citet{2018PASJ...70S...8A} and other relevant papers \citep{2018PASJ...70S...5B,2018PASJ...70...66K,2018PASJ...70S...1M,2018PASJ...70S...2K,2018PASJ...70S...3F}. 
HSC-SSP observations are used due to their wide field of observation, high depth and resolution of its ground-based imaging, giving us access to high quality images of local galaxies. These high quality images are particularly useful for exploting low surface brightness features to identify mergers.

In this work, we use observations from HSC-SSP Wide Public Data Release 3 \citep{2022PASJ...74..247A}. HSC-SSP Wide PDR3  covers about 1100 deg$^2$ of the sky in five broad band filters (\textit{grizy}), with observations from 330 nights coadded, and with a width of $r\approx26$ mags ($5\sigma$, point source), and a seeing resolution of $\sim$0.6--0.8 arcsec. 

We cross-match our HSC-SSP data with spectroscopic observations from Galaxy And Mass Assembly Data Release 4 \citep[GAMA DR4,][]{2022MNRAS.513..439D}, as it gives us access to the galaxy physical properties required for this study. GAMA is essentially magnitude-complete for galaxies brighter than $r=19.8$ mag in the survey area. The cross-matched GAMA-HSC galaxy sample consists of $\sim$144,000 galaxies from $z = 0.01 - 0.35$ with stellar masses $M_* \sim~ 3\times10^9 - 3\times10^{12} M_{\odot}$. 

The next two sections describe \textbf{a)} the selection of mergers in this sample (Section \ref{subsection: mergerselection}) and \textbf{b)} derivation of galaxy physical properties (Section \ref{subsection: galaxyproperties}).
\subsection{Merger Selection}
\label{subsection: mergerselection}
We perform merger classification of HSC-SSP galaxies through deep representation learning, specifically through the use of a fine-tuned version of the publicly available pre-trained model Zoobot \citep{Walmsley2023}. The Zoobot model is pre-trained using data and labels from Galaxy Zoo DECaLS \citep{2022MNRAS.509.3966W}, a project where galaxy images from the deep, low-redshift Dark Energy Camera Legacy Survey \citep[DECaLS,][]{2019AJ....157..168D} are given morphological classifications through citizen science. The use of DECaLS images offers improved imaging quality compared to previous citizen science-based models, and enables for identification of faint, low surface brightness features common in mergers and merger remnants \citep{2019MNRAS.490.5390B,2022scio.confE...2B,2024MNRAS.528.5558W}.

While the Zoobot model can be used as a stand-alone for galaxy morphological classification, in this work we re-train the model for the specific task of merger classification in HSC-SSP using a technique called fine-tuning \citep{goodfellow2016deep}. The fine-tuning technique involves the adaptation of a pre-trained model into a task-specific model through further training using a small sample size. From the initial model, weights, and representations, the uppermost, or `head' layer, is removed, and weights and representations from the remainder of the model frozen. Then, a new `head' model, with outputs tendered for the new task, is added, and the new model is trained for the new task, in the case of this work merger classification within the HSC-SSP. Detailed descriptions of the methods used in GZ DECaLS and Zoobot are available in \citet{2022MNRAS.509.3966W,2022MNRAS.513.1581W,Walmsley2023}.

\begin{figure}[t]
    \centering
    \includegraphics[width=0.4\textwidth]{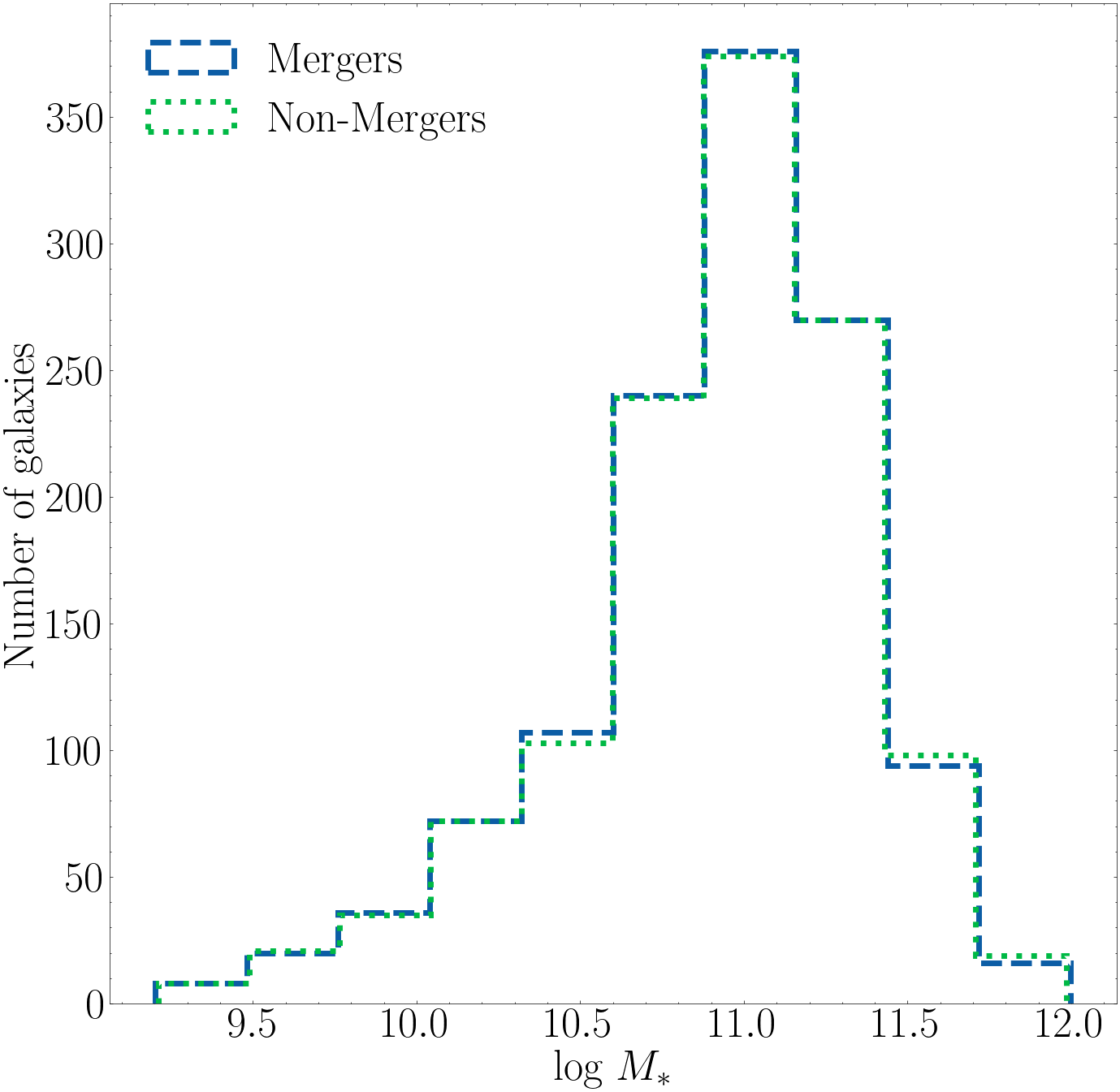}\hfill
    \includegraphics[width=0.4\textwidth]{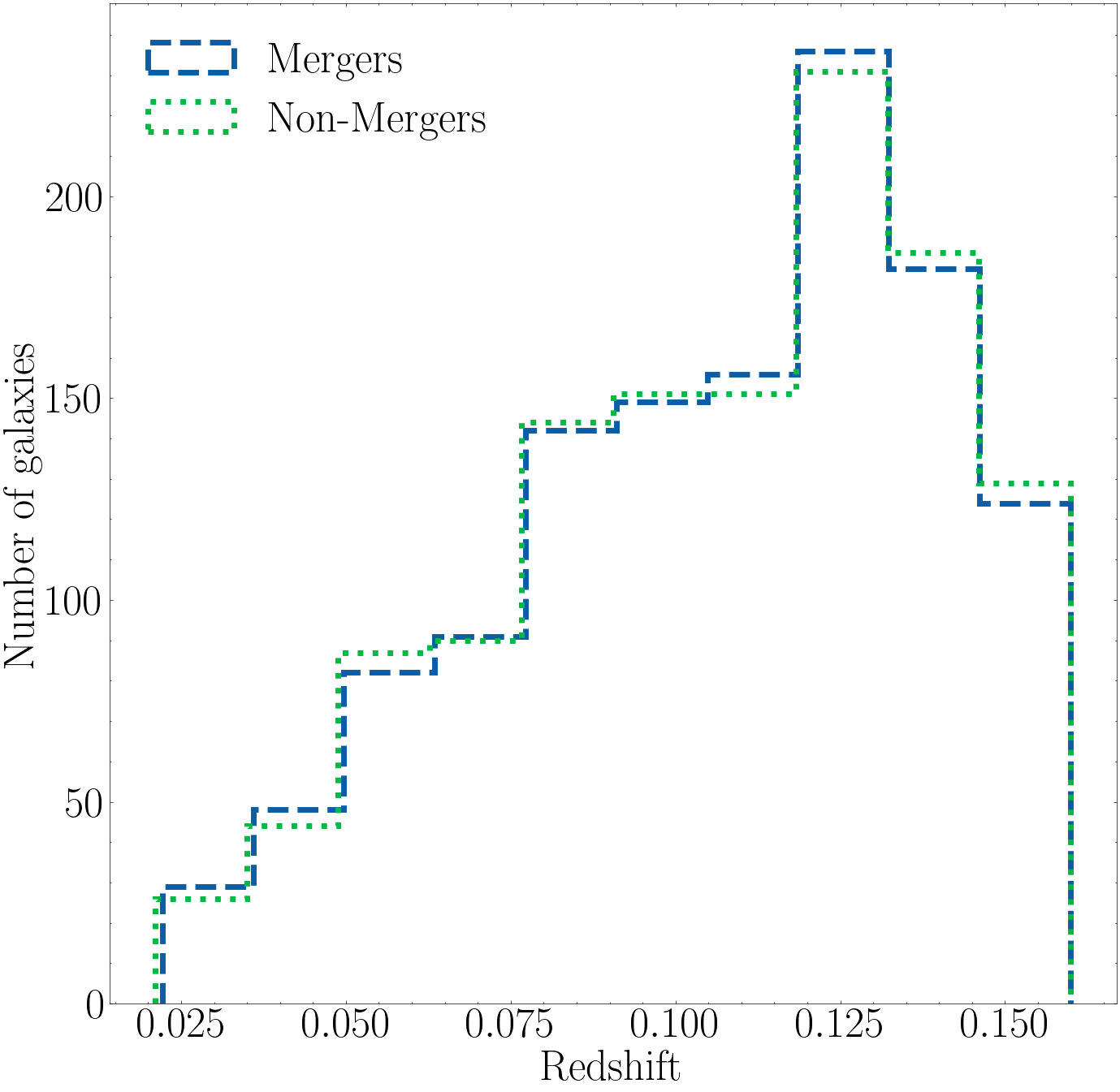}
    \caption{Stellar mass (upper panel) and redshift (lower panel) distributions for the Galaxy Cruise-based merger and non-merger galaxies used for fine-tuning Zoobot. There are $\sim$1200 each of mergers and non-mergers. Each merger galaxy used in the fine-tuning process has a corresponding non-merger galaxy with a similar stellar mass. KS-test results for stellar mass: KS-statistic=0.004, P-value=1.0, for redshift: KS-statistic=0.015, P-value=0.999.
}
    \label{fig:mstars}
\end{figure}

Fine-tuning using Zoobot for merger classification within the HSC-SSP has been done in previous studies \citep{2023A&A...679A.142O}. In \citet{2023A&A...679A.142O}, Zoobot is fine-tuned using mock HSC-SSP images produced from the TNG50 cosmological magneto-hydrodynamical simulation \citep{2019MNRAS.490.3196P, 2019MNRAS.490.3234N} with full details on the image data available in \citet{2024MNRAS.527.6506B}. Since the publication of \citet{2023A&A...679A.142O}, an observation-based HSC-SSP morphology catalog has become publicly available, which is better suited for this work. 
Merger probabilities from Galaxy Cruise \citep{2023PASJ...75..986T} are used to prepare the fine-tuning sample. In Galaxy Cruise, $\sim$20,000 galaxies from HSC-SSP are given visual inspection-based merger probabilities through citizen science based on over 2 million classifications. To obtain merger probabilities, first an initial merger probability $P(int.)$ is evaluated by dividing the number of merger classifications by the number of total classifications ${N_{int.}}/{N_{total}}$. Next, each human classifier is assigned a $P(bad)$ value, or a probability that a classifier makes an inaccurate classification, calculated based on how often the classifier made a classification disagreeing with the majority. Then, $P(int.)$ is re-calculated removing any classifications made by classifiers with $P(bad) > 0.1$, which reduces the risk of misclassified galaxies. The fine-tuning sample is prepared based on these merger probabilities. Mergers, assigned a training label of 1, are all galaxies in the Galaxy Cruise catalog with $P(int.) > 0.79$. Non-mergers, assigned a training label of 0, are selected from galaxies with $P(int.) < 0.3$, with the expectation that such a low threshold will feature very few galaxies with visual merger features, thereby decreasing the risk of contamination in the non-merger sample. The $P(int.) > 0.79$ cut gives us $\sim$1200 mergers, and for each merger we selected a non-merger with similar stellar mass and redshift. We show the distribution of stellar mass and redshifts of our fine-tuning sample for each class in Fig. \ref{fig:mstars}, with KS-test results available in the captions. We fine-tune Zoobot using \textit{gri} images of these $\sim$2400 galaxies, each first cutout to a size encompassing $10\times R_{\textrm{eff}}$, then re-sized to 300 $\times$ 300 pixels. Fine-tuning with a sample size of this order has been shown to be sufficient for merger classification in \citet{2023A&A...679A.142O}. The fine-tuning procedure returns an accuracy of $83\%$, a recall of $84\%$ and precision of $84\%$ on the fine-tuning sample. These numbers are similar to \citet{2023A&A...679A.142O}, but it should be noted that a perfect accuracy is not expected in machine-learning based galaxy morphological classifications due to the difficulty of the task \citep{2021MNRAS.504..372B,2022MNRAS.511..100B,2023MNRAS.521.3861D}.

\begin{figure}[h!]

    \centering
    \includegraphics[width=0.5\textwidth]{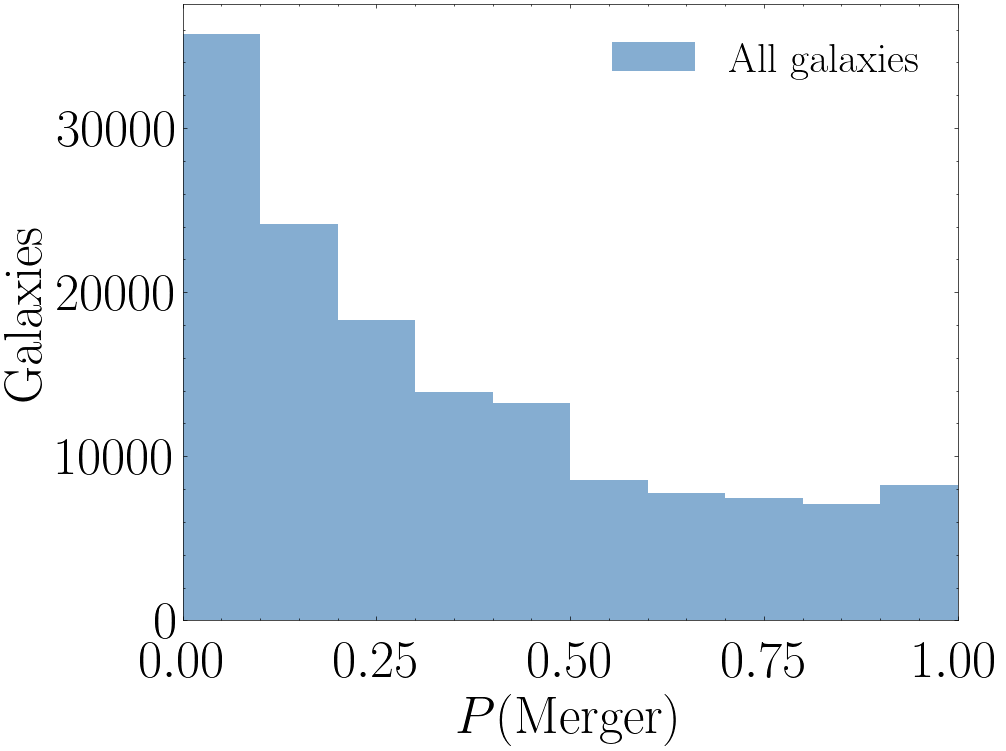}
    \caption{Merger probability distributions for HSC-GAMA cross-matched galaxies predicted using the Zoobot model fine-tuned using Galaxy Cruise images and labels. We find that the most galaxies have very low merger probabilities, and the number of galaxies in each probability bin decreases as probability increases.}
    \label{fig:HSCGAMA}
\end{figure}

After fine-tuning is complete, predictions are made on \textit{gri} images of HSC-SSP galaxies. Similar to the fine-tuning data, the predicted images are cutout first with a size encompassing $10\times R_{\textrm{eff}}$, then re-sized to 300 $\times$ 300 pixels. For each input image, our model will output a merger probability between 0 and 1, with 0 indicating full confidence in a non-merger and 1 indicating full confidence in a merger. We show the probability distribution of the merger probability in Fig. \ref{fig:HSCGAMA}. We find that the most galaxies ($\sim$78000) are given merger probabilities below 0.3, and the number of galaxies in each probability bin decreases as probability increases, with $\sim$15000 galaxies being given a merger probability $>0.8$. This distribution greatly differs from predictions given in the simulation-based network in \citet{2023A&A...679A.142O}, where the peak of the distribution was in the unclear merger probability range (0.4 - 0.6), even if the probabilities were assigned on a very similar set of galaxies (HSC-SSP cross-match with GAMA). One possible reason is that \citet{2023A&A...679A.142O} uses mini mergers with merger mass ratio $< 1:20$ to fine-tune Zoobot. Such galaxies are visually difficult to differentiate from non-mergers, which may lead to unconfident predictions made by the model.

\subsection{Galaxy Properties}
\label{subsection: galaxyproperties}
We use stellar mass, star formation rates, and AGN properties for GAMA galaxies from the table produced by \citet{2022MNRAS.509.4940T}. The properties are derived using the SED fitting code \textsc{ProSpect} \citep{2020MNRAS.495..905R}. \textsc{ProSpect} fits across a broad range of wavelengths, ranging from the far-ultraviolet (FUV) to far-infrared (FIR). These bands are GALEX \textit{FUV} and \textit{NUV} \citep{2007ApJS..172..468Z}; VST \textit{u, g, r, i} \citep{2021Msngr.183...25I}; VISTA \textit{Z, Y, J, H, $K_{\mathrm{S}}$} \citep{2012A&A...544A.156M}; \textit{WISE W1, W2, W3, W4} \citep{2012yCat.2311....0C}; and \textit{Herschel} P100, P160, S250, S350, and S500 \citep{2011A&A...532A..90L,2012MNRAS.424.1614O}. Further details on the derivations of the photometric bands in GAMA DR4 are available in \citet{2018MNRAS.476.3137R,2020MNRAS.496.3235B}. Photometry compiled by \citet{2021MNRAS.506..256D} is put through \textsc{ProSpect}, which uses a \citet{2003MNRAS.344.1000B} stellar library and \citet{2003PASP..115..763C} initial mass function to model stellar components. The model consists of a two-component description of the interstellar medium, a dust component attenuating emission from all stars, and a birth cloud attenuating emission from young stars and the birth cloud itself. Details about \textsc{ProSpect} are available in \citet{2020MNRAS.495..905R}, and further details regarding galaxy modeling are available in \citet{2021MNRAS.505..540T,2022MNRAS.509.4940T}, but we will summarize the main points here.

The star formation rates and stellar masses follow \citet{2021MNRAS.505..540T}. To obtain these properties, a parametrized star formation history (SFH) is adopted. The star formation model takes the form of a skewed normal distribution in lookback time. The peak position of star formation, peak SFR, SFH width, and skewness are set as free parameters in this model. This model allows for the recovery of a variety of SFHs. 

A major characteristic of the \textsc{ProSpect} fitting code is its ability to simultaneously model the contributions from the galaxy and AGN components, and distinguish the flux contribution between the two. The \citet{2006MNRAS.366..767F} AGN model is used by \textsc{ProSpect}, as it can fit MIR excess and the larger wavelength coverage in the ultraviolet, allowing for greater constraints on AGN parameters in tandem with star formation. The Fritz model is able to account sufficiently for the contribution from the galaxy as well as from the AGN torus. Various AGN implementations and their contributions to the overall galaxy SED can be accounted for by the Fritz model. The significance of simultaneous modeling is that it is able to identify AGNs masking as star formation. A model that only accounts for the galaxy component can overestimate stellar masses and SFRs in the presence of an AGN, as the AGN light could be erroneously interpreted as stellar or nebular light \citep{2023ApJ...959L..18D}.

We quantify AGN activity based on two values in the catalogue recovered by \textsc{ProSpect}, provided by \citet{2022MNRAS.509.4940T}. The first value is the AGN bolometric luminosity in units of solar luminosity $L_{\mathrm{AGN}}$. The second value is $f_{\mathrm{AGN}}$ value, which is the fraction of flux within $5 - 20 \mu $m contributed by the AGN component relative to the entire flux in the wavelength range. The reason we include $L_{\mathrm{AGN}}$ in our investigation in addition to $f_{\mathrm{AGN}}$, is because $L_{\mathrm{AGN}}$ is likely to hold information about the accretion of the AGN independent of the galaxy flux. That is, if an AGN host also has high SFR or stellar emission, it may result in a decreased $f_{\mathrm{AGN}}$. While previous works have investigated the connection between mergers and AGN bolometric luminosities \citep{2012ApJ...758L..39T,2018MNRAS.480.3562D}, this is the first work that conducts merger-AGN connections that identifies AGNs using a unified model that simultaneously obtains galaxy and AGN properties. We note that we remove any galaxies with $L_{\mathrm{AGN}} < 10^{36} L_{\odot}$ from our investigations. At low luminosities, the galaxy emission is difficult to differentiate from the AGN component, and \textsc{ProSpect} cannot recover a well-constrained AGN luminosity.

\subsection{Visual Purification of Merger Candidates}

Our model is unable to assign probabilities or classify between merger stages, as it is a pure binary classifier, or in other words, our model is only able to assign a probability on how merger-like morphology a galaxy has. As such, we follow a similar approach to \citet{2022MNRAS.514.3294B}, where machine learning methodology is used for initial selection of merger candidates, then followed up with visual classification. For this work, we conducted a visual followup on 3000 of the $\sim7000$ merger candidates, or galaxies given a merger probability $>0.8$ by our model, to visually identify pair stage mergers and postmergers (2000 by KO, 1000 by CB). We choose this threshold as it is very similar to the cutoff used to select mergers in the Galaxy Cruise catalog. The followup was conducted using HSCMap \footnote{\url{https://hscmap.mtk.nao.ac.jp/hscMap4/}} in conjunction with the same images as the ones the model made classifications on, being \textit{gri}-band three color HSC images cutout with a size encompassing $10\times R_{\textrm{eff}}$, then re-sized to 300 $\times$ 300 pixels. 

postmergers were required to be fully coalesced, that is, the galaxy only has one distinguishable nucleus. In addition, postmergers were required to have visual signatures attributed to merger activity, such as rings, shells, or streams, or other extended or asymmetric features. For pairs, we required that the galaxy has two distinguishable nuclei. In addition, we required that the galaxy pair has clear visual morphological merger signatures, such as tidal tails or bridges between nuclei. The requirement for distinguishable visual merger signatures limits contamination from early stage mergers, which would likely be before the stage of any star formation or AGN activity, from our pair sample. We also conduct spectroscopic confirmation on the pairs where information is available for both sets of the pair, and consider the galaxies within 50 kpc and $\Delta v =500$ km/s of each other as close pairs. KO and CB reviewed each other's followups to ensure consistent criteria for pairs and postmergers. These visual follow-ups resulted in a sample of 259 pairs and 100 post mergers. Of the pairs that had spectroscopic observations available for both galaxies, we considered 16 to be close pairs, being $50 \mathrm{kpc} h^{-1}$ of each other and $\Delta v < 500 \mathrm{km/s}$. Mosaics for merger candidates, control candidates, pair sample, and postmerger sample are available in Appendix \ref{appendix: A}, with the complete figure set available in the online journal. Each mosaic has 20 galaxies, with merger probabilities in decreasing order.

\subsection{Selection of SFGs}
\label{subsection:SFGselect}
\begin{figure}[h!]

    \centering
    \includegraphics[width=0.5\textwidth]{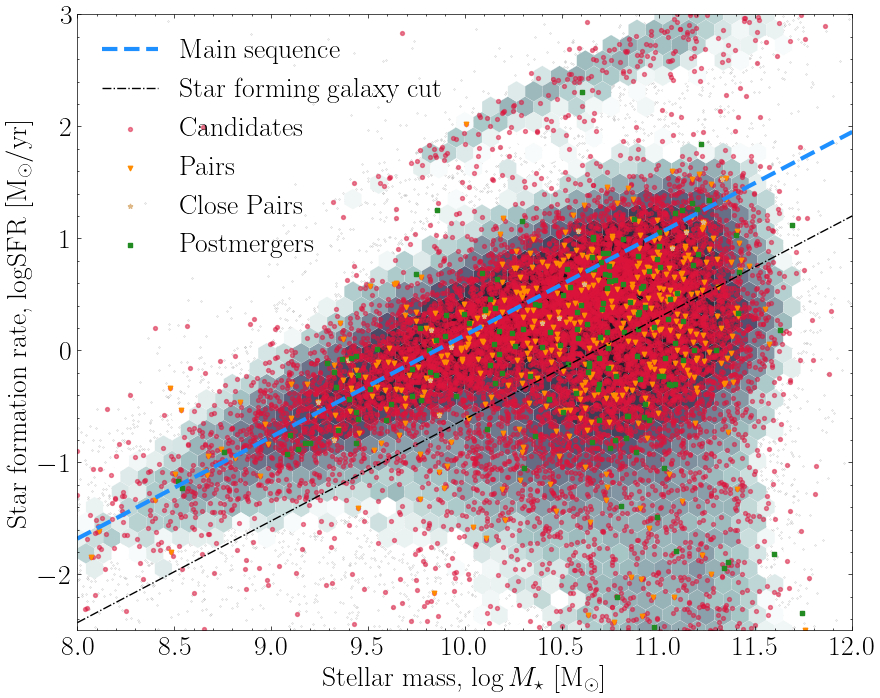}
    \caption{Selection of SFGs from the HSC-SSP - GAMA cross-matched sample. The entire sample is plotted in blue. The main sequence found through the iterative approach in \citet{2019MNRAS.485.4817D, 2021MNRAS.506.4760D} is indicated by the solid line. Galaxies within 0.75 dex below the main sequence (dash-dotted line) are used in our studies. Galaxies outside of this limit, as well as those with LP$<-30$, poorly constrained SFRs ($\mathrm{SFR error} > \mathrm{SFR}/2$) or unreliable AGN lumminosities ($L_{\mathrm{AGN}}<10^{36}L_\odot$)} are removed from any studies. We also indicate the \textbf{a)} merger candidates (merger probability $>$ 0.8, red dots), \textbf{b)} visually confirmed pairs (orange triangles), \textbf{c)} visually and spectroscopically verified close pairs (beige stars), and d) postmergers (green squares).
    \label{fig:ms}
\end{figure}

Figure \ref{fig:ms} shows the stellar mass - SFR relation of all of our galaxies. As this part of this work investigates the quantitative connection between merger activity and star formation, we limit both our merger and control samples to starforming galaxies (SFGs). To select SFGs, we follow an iterative approach laid out in \citet{2019MNRAS.485.4817D, 2021MNRAS.506.4760D}. In each iteration, the median star formation rates (SFRs) of galaxies as a function of galaxy stellar mass are computed to $M_* < 10^{10.2} M_{\odot}$ in logarithmic stellar mass bins of 0.2 dex. Galaxies with SFR lower than 1 dex below the median SFR are considered passive, and removed from subsequent iterations of median calculation. This process is iterated until the median in each bin converges to within 1\%. The midpoints for each mass bin and stellar mass medians are then fit to give us a main sequence which is extrapolated at higher masses. Galaxies with SFRs less than below 0.75 dex the final main sequence are considered as quenched and are removed from any subsequent investigations. We follow \citet{2020MNRAS.493.3716H} with the removal of quenched galaxies, stressing the importance of class matching. Without these cuts between SFG and quenched galaxies, a SFG merger could be given an unfair control that is quenched, and vice versa. We note that before the iterative approach, we remove galaxies that are likely to have poorly constrained physical properties which make them unsuitable for science. These are galaxies where \textsc{ProSpect} returns a log-posterior (LP) value of LP$<-30$, galaxies with poorly constrained SFRs (SFR errors $ > \mathrm{SFR}/2$), or galaxies and galaxies with unreliable AGN luminosities ($L_{\mathrm{AGN}}<10^{36} L_\odot$).  We show our selection of SFGs in Fig. \ref{fig:ms}. The SFG selection and removal of unreliable galaxies left us with $\sim7000$ merger candidates with merger probability $>0.8$.

\subsection{Control Matching Procedures}
For each galaxy we use in the investigation in the following section, we match 50 control galaxies (merger probability  $< 0.3$) in the stellar mass and redshift space. This search was with replacement, that is, a single control galaxy can be a control for multiple mergers.
As some controls may be closer matches than others, we adopt the statistical weighting method done in \citet{2016MNRAS.461.2589P}. First, we search for matches within the tolerances for each property, initially set to $\Delta \log M_{*} < 0.01$ dex and $\Delta z < 1$ percent. If a control is not found within the set tolerances, the tolerances are increased by 0.01 dex for stellar mass and one percent for redshift, up to a maximum of five increases. Galaxies that do not have 50 controls are removed from the investigation.
Next, a weighting scheme is applied so that the closest matches are assigned larger statistical weights. A galaxy with identical mass and redshift would be assigned a statistical weight of one, while the worst matches, that is, galaxies with physical properties at the limits of tolerance, are assigned a weight of zero. For example, for a merger galaxy with stellar mass $M$ and stellar mass tolerance $M_{\mathrm{tol}}$, the stellar mass weight $w_{M,i}$ for the $i$th control galaxy with stellar mass $M_i$ is obtained through

\begin{equation}
     w_{M,i} = 1-\frac{|M-M_{i}|}{M_{\mathrm{tol}}}.
\end{equation}
   
The redshift statistical weight $w_{z,i}$ is calculated similarly. The overall statistical weight for the $i$th control galaxy is given by

\begin{equation}
    w_{i} = w_{M,i}w_{z,i}.
\end{equation}

The statistical weight can then be used to calculate the weighted mean for any physical property $x$ for a control set of a given merger galaxy. For a galaxy with $N$ controls, the weighted mean $x_{\mathrm{mean}}$ for physical property $x$ can be obtained by

\begin{equation}
    <x> = \frac{\Sigma_{i=1}^N x_i w_i}{\Sigma_{i=1}^N w_i}.
\end{equation}

Any galaxies where 50 controls were not found were removed from any subsequent analysis. Table \ref{table:1} summarizes the number of galaxies used.

\begin{table}[h!]
\begin{tabular}{|l c c|}
\hline
\textbf{Classification}&\vtop{\hbox{\strut \textbf{Initial}}\hbox{\strut \textbf{sample}}}&\vtop{\hbox{\strut \textbf{Final}}\hbox{\strut \textbf{sample}}}\\
\hline
\vtop{\hbox{\strut Merger candidates}\hbox{\strut (P(merger)$>0.8$)}}&6803&6745\\
\hline
\vtop{\hbox{\strut Random controls}\hbox{\strut  (P(merger)$<0.3$)}}&1000&982\\
\hline
\vtop{\hbox{\strut Visually confirmed}\hbox{\strut pairs}}&259&249\\
\hline
Close pairs&16&15\\
\hline
\vtop{\hbox{\strut Visually confirmed}\hbox{\strut postmergers}}&100&98\\
\hline

\end{tabular}
\caption{A summary of the number of galaxies of each classification used in this study. Initial sample points to the number of galaxies after CNN classification, visual purification, and SFG selection. Final sample points to the number of galaxies after control matching, with galaxies without 50 controls removed.}
\label{table:1}
\end{table}

\section{Results}
\label{section:Results}

We conducted multiple tests on the connection between mergers, star formation, and AGN activity, on five samples. The first sample is the entire set of merger candidates selected based on the output of our fine-tuned model, being galaxies with merger probability $\geq$ 0.8. The second and third samples are the visually purified pairs and postmergers, respectively. The fourth sample is a randomly selected set of 1000 pseudo-mergers, selected from the control sample of galaxies with merger probability $<$ 0.3. This dataset is to validate that the enhancements are not due to stochasticity in the merger samples. The fifth sample is also a validation dataset, containing all galaxies with merger probability $>$ 0.3. Each merger in each sample is matched to 50 non-mergers using the method highlighted in the previous section. The AGN properties $f_{\mathrm{AGN}}$ and $L_{\mathrm{AGN}}$ from \citet{2022MNRAS.509.4940T} are used as parameters to quantify AGN activity. For each merger, we find $\Delta$ SFR, $\Delta f_{\mathrm{AGN}}$, and $\Delta L_{\mathrm{AGN}}$ by comparing the weighted mean of these values of the controls from those of the merger. Errors are propagated from the 16th and 84th percentiles from the \textsc{ProSpect} measurement posteriors. For example, to calculate the error in SFR, we use $(\mathrm{SFR}_{84}-\mathrm{SFR}_{16})\times0.5$, with the 16th and 84th percentile values returned by \textsc{ProSpect}. In this section, there is an `enhancement' in SFR, $f_{\mathrm{AGN}}$, and $ L_{\mathrm{AGN}}$ if $\Delta$ SFR, $\Delta f_{\mathrm{AGN}}$, and $\Delta L_{\mathrm{AGN}}$, respectively are greater than 0, indicating that the median value for the property in the merger sample is greater relative the controls. Conversely, a `suppression' is when the value is less than 0, or the median value for the property is greater in the controls than the merger samples.

\subsection{SFR Enhancement}
First, we investigated whether there existed enhancements in star formation activity in mergers with respect to their control galaxies. The SFR enhancement of each galaxy relative to each of their \textit{N} of controls is obtained through
\begin{equation}
    \Delta\mathrm{SFR}=\frac{\Sigma_{i=1}^N\log({\mathrm{SFR_{merger}}}/{\mathrm{SFR_{control}i}}) w_i}{\Sigma_{i=1}^N w_i}
\end{equation}
Figure \ref{fig:sfrenhancement} shows the mean SFR enhancement for each dataset. We find that the entire merger candidate sample shows a meager suppression of SFR relative to their controls, $\Delta\mathrm{SFR}=-0.009\pm0.003$ dex. The pair sample also shows meager suppression, $\Delta\mathrm{SFR}=-0.033\pm0.014$ dex, and the close pair sample shows an enhancement of $\Delta\mathrm{SFR}=0.013\pm0.064$ dex. The postmerger sample is suppressed, with $\Delta\mathrm{SFR}=-0.057\pm0.024$ dex.

\subsection{\texorpdfstring{$f_{\mathrm{AGN}}$}{fAGN} Enhancement}
 Next, we investigate whether there exists enhancement in the AGN flux contribution, $f_{\mathrm{AGN}}$, between mergers and their respective controls. The AGN enhancement enhancement of each galaxy relative to each of their \textit{N} controls is obtained through
\begin{equation}
    \Delta f_\mathrm{AGN}=\frac{\Sigma_{i=1}^N(\log(f_\mathrm{AGN_{merger}}/f_\mathrm{AGN_{control}i}) w_i}{\Sigma_{i=1}^N w_i}
\end{equation}
Figure \ref{fig:fagnenhancement} shows the mean $f_{\mathrm{AGN}}$ enhancement for each sample. $f_{\mathrm{AGN}}$ is enhanced in the pair and postmerger samples, with the enhancement greater in the postmerger sample. The entire merger candidate sample shows a suppression relative to controls, $\Delta f_{\mathrm{AGN}}=-0.010\pm0.033$ dex. The pair sample shows a suppression in $f_{\mathrm{AGN}}$ relative to their controls, $\Delta f_{\mathrm{AGN}}=-0.024\pm0.170$ dex, and the close pair sample shows an enhancement of $\Delta f_{\mathrm{AGN}}=0.124\pm0.779$ dex. The postmerger sample is also enhanced in $f_{\mathrm{AGN}}$, with $\Delta f_{\mathrm{AGN}}=0.286\pm0.270$ dex.

\subsection{\texorpdfstring{$L_{\mathrm{AGN}}$}{LAGN} Enhancement}
Third, we investigate whether there exists enhancement in AGN luminosity, $L_{\mathrm{AGN}}$ between mergers and their respective controls. The AGN luminosity enhancement of each galaxy relative to each of their \textit{N} controls is obtained through
\begin{equation}
    \Delta L_\mathrm{AGN}=\frac{\Sigma_{i=1}^N(\log(L_\mathrm{AGN_{merger}})-\log(L_\mathrm{AGN_{control}i})) w_i}{\Sigma_{i=1}^N w_i}
\end{equation}
Figure \ref{fig:Lagnenhancement} shows the mean $L_{\mathrm{AGN}}$ enhancement of each sample. Similar to the $f_{\mathrm{AGN}}$ enhancement results, while there exists an enhancement in $L_{\mathrm{AGN}}$ in the pair and postmerger samples, the average enhancement is greatest in the postmerger sample compared to the pair sample. The entire merger candidate sample shows an enhancement in $L_{\mathrm{AGN}}$ compared to their controls, $\Delta L_{\mathrm{AGN}}=0.002\pm0.025$ dex. The pair sample shows an enhancement of $\Delta L_{\mathrm{AGN}}=0.019\pm0.129$ dex, and the close pair sample shows enhancement of $\Delta L_{\mathrm{AGN}} = 0.127\pm0.461$ dex. The postmerger sample is shows an enhancement of $L_{\mathrm{AGN}}$ compared to their controls of $\Delta L_{\mathrm{AGN}} = 0.329\pm0.195$ dex.

\begin{figure}[h!]
    \centering
    \includegraphics[width=0.5\textwidth]{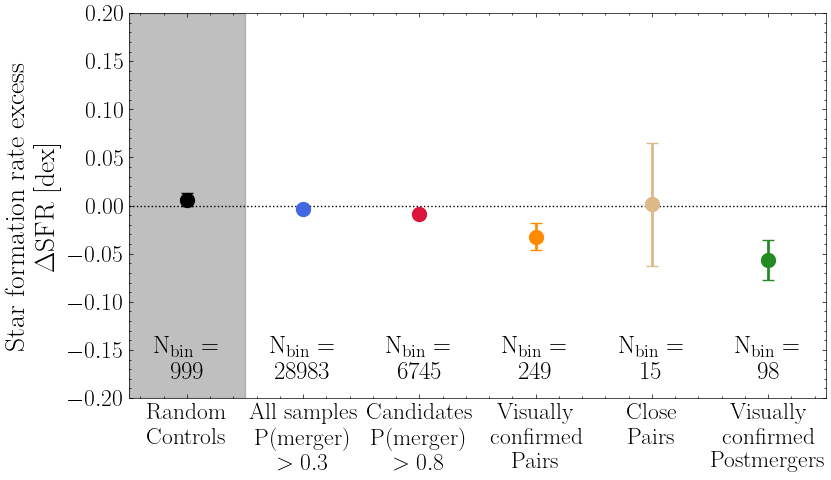}
    \caption{Average star formation enhancements for each merger sample. From left to right, the datapoints indicates the mean $\Delta $SFR, or mean star formation enhancement, for \textbf{a)} the `fake' merger control sample (black points with grey background), \textbf{b)} all galaxies with merger probability $>0.3$ (blue points) \textbf{c)} the entire merger sample (merger probability $>0.8$) before visual purification (red points), \textbf{d)} the visually purified pair sample (orange points), \textbf{e)} of the visually purified pairs, those which are close pairs, and \textbf{f)} the visually purified postmerger sample (green points). The number of galaxies in each sample are indicated below each point. The error bars indicate standard errors based on the 16th and 84th percentile measures obtained by \textsc{ProSpect}. The horizontal black dashed line indicates 0, or no enhancement. We find that any enhancement or suppression existing between the merger and control samples are very small, for all samples.}
    \label{fig:sfrenhancement}
\end{figure}

\begin{figure}[h!]
    \centering
    \includegraphics[width=0.5\textwidth]{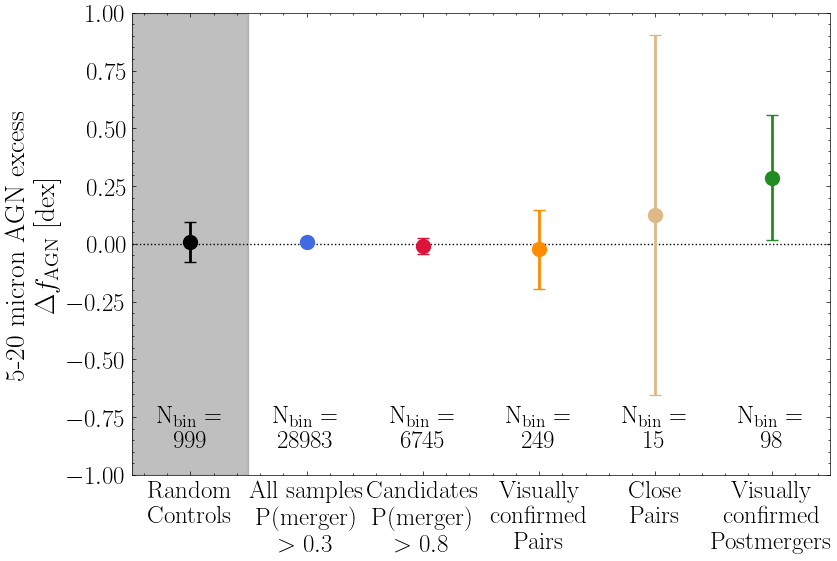}
    \caption{The same as Fig. \ref{fig:sfrenhancement}, but for $f_{\mathrm{AGN}}$ enhancement. As there are no percentile measures available for $f_{\mathrm{AGN}}$, the error bars indicate uncertainties. The number of mergers in each sample is indicated below each sample. We find that the $f_{\mathrm{AGN}}$ enhancement is greatest among the close pair and postmerger samples.}
    \label{fig:fagnenhancement}
\end{figure}

\begin{figure}[h!]
    \centering
    \includegraphics[width=0.5\textwidth]{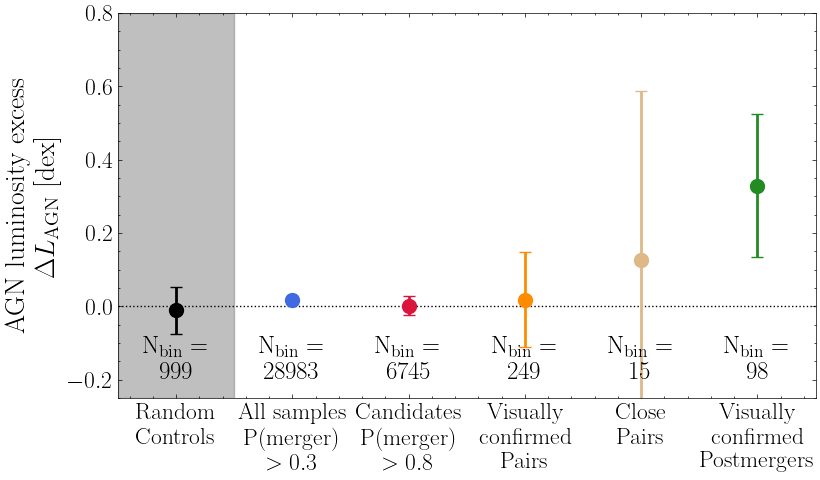}
    \caption{The same as Fig. \ref{fig:sfrenhancement}, but for $L_{\mathrm{AGN}}$ enhancement. Similar to $f_{\mathrm{AGN}}$ enhancement, we find the enhancement being greatest among the close pair and postmerger samples.}
    \label{fig:Lagnenhancement}
\end{figure}

We note that both the average enhancements and standard errors are sensitive to the number of controls used. We make available each of these figures with five, 20, and 100 controls in  Appendix \ref{appendix: B} as reference.

\subsection{AGN Excess}

\begin{figure}[h!]
    \centering
    \includegraphics[width=0.5\textwidth]{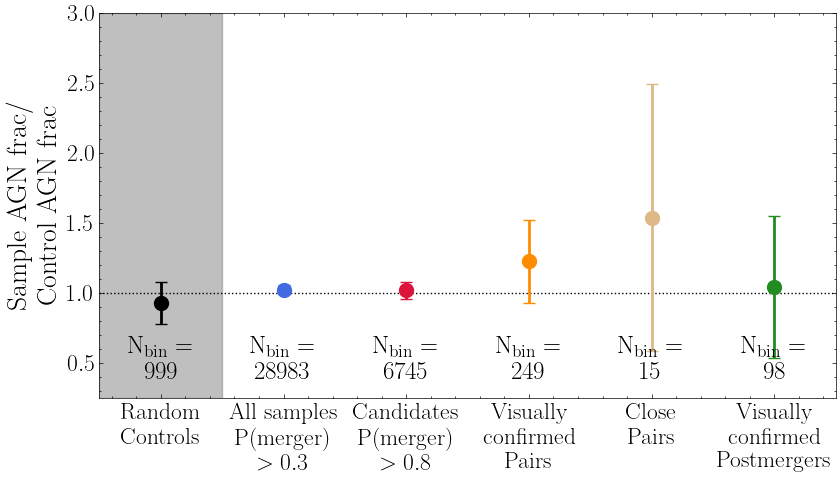}
    \caption{The AGN excess, defined as the fraction of mergers that host AGNs over the fraction of controls that host AGNs, for each sample. The horizontal dashed line indicates a ratio of 1, which means no difference between AGN host fraction between mergers and controls. We find the greatest AGN excess to be among pairs and postmergers.}
    \label{fig:agnexcess}
\end{figure}

In addition to the enhancements in $f_{\mathrm{AGN}}$ and $f_{\mathrm{AGN}}$, we investigate the AGN excess in mergers. Figure \ref{fig:agnexcess} shows the AGN excess among each sample. AGN excess is the fraction of AGN hosts among each sample over the fraction of AGN hosts among the controls, or

\begin{equation}
    \frac{n_{\mathrm{AGN},\mathrm{merger}}/n_\mathrm{merger}}{n_{\mathrm{AGN},\mathrm{control}}/n_\mathrm{control}}.
\end{equation}

To identify AGN hosts, we follow \citet{2022MNRAS.509.4940T}. In \citet{2022MNRAS.509.4940T}, AGN hosts are identified as galaxies where \textsc{ProSpect} returns  $f_{\mathrm{AGN}}>0.1$. An AGN excess of 1 finds the same AGN fraction in the merger sample relative to its controls. We find that while the entire merger candidate sample finds little AGN excess, at $1.021\pm0.061$, AGN excess is found among the pair and postmerger stages, with $1.225\pm0.296$ for pairs and $1.538\pm0.954$ for close pairs, and $1.104\pm0.509$ for postmergers.

\section{Discussion}
\label{section:Discussion}

With regard to star formation, we find very small differences between the mergers and controls for the merger candidate and pair samples, an enhancement in the close pair sample and a suppression in the postmerger sample.
With regard to AGN activity, while the entire merger candidate sample does not have notable enhancement in both $f_{\mathrm{AGN}}$ and $L_{\mathrm{AGN}}$, we find pronounced enhancements among the visually purified postmerger sample.

\subsection{Star formation}
We do not find an enhancement or suppression in star formation in our merger candidate sample. We also do not find any star formation enhancement or suppression in the pair stage, while other studies using spectroscopically confirmed pairs robustly demonstrate enhancements in star formation using emission line diagnostics \citep{2008AJ....135.1877E}. In the postmerger stage, we find a suppression in star formation, while other studies using emission line-based star formation rates have found an enhancement in star formation \citep{2022MNRAS.514.3294B,2024arXiv241006356F,2024arXiv241006357E}. 

\subsubsection{Importance of timescales}
Our disagreements with previous studies may be due to a number of reasons. First, our control sample is likely to include star-forming galaxies which are undergoing active star formation from secular processes unrelated to mergers \citep{2006RMxAC..26..131C}. However, more notably, the disagreements, particularly the suppressions in the postmerger stage, may be a result of the timescale between mergers, star formation, and quenching, and the visual morphological timescales of mergers.

The timescale in which the morphological signatures of postmergers remains observable can be up to 2 Gyr after final coalescence, at surface brightness limits equivalent to HSC-SSP \citep{2008MNRAS.391.1137L,2014A&A...566A..97J,2022MNRAS.511..100B,2024MNRAS.528.5558W}. As such, our postmerger selection may be based on visual features which last on a longer timescale than those of the star formation tracers we use. At this time, while the visual features have not yet dissipated, the merger induced starburst may cease, and the subsequent consumption of gas and feedback resulting from star formation and AGN activity may cause the system to go to the green valley or even quench
\citep{2015MNRAS.452..616D,2022MNRAS.515.1430D,2025OJAp....8E..12E}. These potentially quenched systems may have very low SFRs if instantaneous star formation indicators such as emission lines are used, but appear to be SFGs based on SFRs recovered by \textsc{ProSpect}. These SFRs are time-averaged over approximately 100 Myr \citep{2020MNRAS.495..905R} as opposed to a shorter timescale indicator such as emission lines, which trace up to approximately 10 Myr \citep{2022MNRAS.513.2904T}. As such, SFRs recovered by \textsc{ProSpect} may be an average of the potential starburst and quench phase of the merger process, which can take place rapidly \citep{2024OJAp....7E.121E}.

\subsubsection{Primaries and secondaries in pairs}
With regard to the lack of enhancement in pairs, in addition to the difference in timescales, the pair sample may include pairs without a pronounced increase in star formation, such as minor merger pairs \citep{2023RAA....23i5026D} or dry mergers \citep{2006ApJ...636L..81N}. Also, the enhancement of SFR in galaxy pairs has been shown to differ between the primary and secondary galaxy of the pair \citep{2013MNRAS.431..167R,2015MNRAS.452..616D}. Specifically, SFR enhancements are reported in the primary while suppressions are reported in the secondary. This behavior may arise as a consequence of the secondary being cut off from accretion or tidally stripped of extragalactic gas in the larger halo of the primary. We may have a collection of both primary and secondary galaxies in our pair sample, resulting in the enhancements and suppressions averaging out to zero. We note that we only classify pairs between all pairs and close pairs, and no additional classification is made such as mass ratio or the morphology of the galaxies in the pair, which have been shown to affect the star formation enhancement among pairs \citep{2007A&A...468...61D,2022MNRAS.514.3294B,2024ApJ...965...60F}. 
A further decomposition of the pair sample into properties such as mass ratio and pair morphology may yield further differences in enhancement between sub-samples.

However, most notably, similar to the postmerger sample, the lack of noticeable enhancement or suppression among any sample is likely due to the SFRs recovered by \textsc{ProSpect} being time-averaged over 100 Myr \citep{2022MNRAS.513.2985R} . While star formation enhancements happen in the close pair stage of the merger process, the SFRs recovered by \textsc{ProSpect} likely probes the stages both prior to and after the star formation enhancement, averaging out to no enhancement or suppression. These results are consistent with \citet{2015MNRAS.452..616D}, which finds no enhancement or suppression among major or minor mergers in both the primary and secondary galaxy for longer time-scale SFR indicators. In addition, for previous works, there exists a possibility that SFR is over-estimated in presence of an AGN, resulting in the SFR enhancements being more pronounced compared to our results.

\subsection{AGN activity}
Regarding AGN activity, our results find an enhancement among close pairs and postmergers, but no enhancement in the other samples. Previous works such as \citet{2012ApJ...758L..39T}, \citet{2012A&A...540A.109S}, \citet{2015ApJ...814..104K}, \citet{2018PASJ...70S..37G}, \citet{2019MNRAS.487.2491E} find an increase in AGN activity among mergers, but others such as \citet{2005ApJ...627L..97G}, \citet{2009ApJ...691..705G}, \citet{2011ApJ...726...57C}, \citet{2012ApJ...744..148K}, \citet{2014MNRAS.439.3342V}, \citet{2016ApJ...830..156M}, \citet{2020A&A...637A..94G}, \citet{2021ApJ...909..124S} question the role of mergers in AGN triggering activity. The mixed results compared with previous studies may be due to a number of reasons.
First, similar to our results for SFR, the control sample may have elevated AGN activity due to secular processes unrelated to mergers. In particular, at low redshifts ($z<1$), secular processes are the primary mechanism for AGN triggering compared to mergers \citep{2012ApJ...751...72D, 2019MNRAS.489.4016S}. In addition, the discrepancy may also be due to the diversity of the full merger sample. The fine-tuned model used in this work is able to classify merger galaxies with diverse appearances, and does not discriminate between merger stage or merger properties.
The galaxy merger process can last up to several Gyr progressing through all of the stages \citep{2008MNRAS.391.1137L,2010MNRAS.404..575L,2010MNRAS.404..590L}.
In comparison, the lifetime of an AGN is typically much shorter, ranging from days to $10^{7-8}$ years \citep{1996ApJ...470..322C,1996ASPC..110....3S,2004MNRAS.351..169M,2015MNRAS.451.2517S,2021Sci...373..789B}. As such, it is likely that while mergers can trigger AGN activity, many of the mergers selected by our methods are not at a stage where enhanced AGN activity can be observed, such as early stage pairs or pre-coalescence.
Also, the diverse appearances of mergers selected by the model means that mergers with diverse physical properties are given higher merger probabilities, suggesting that mergers that are less likely to induce AGNs, such as gas poor dry mergers \citep{2021ApJ...909..124S}, are included in the full merger candidate sample, lowering the average enhancement. In addition, many of the previously mentioned studies find that a merger-AGN connection are focused on major mergers, so limiting our merger sample to major mergers may alter our results, even though the connection between merger and AGN activity for gas-rich major mergers has also been questioned \citep{2019ApJ...882..141M}. Indeed, in our visually purified samples, the enhancement in $f_{\mathrm{AGN}}$ and $L_{\mathrm{AGN}}$ are very small in the pair stage, which is consistent with works such as \citet{2000ApJ...530..660B,2007MNRAS.375.1017A,2008AJ....135.1877E}, suggesting that merger-induced AGN ignition is delayed relative to the pair stage. For visually purified postmergers, there is an  increase in both $f_{\mathrm{AGN}}$ and $L_{\mathrm{AGN}}$ enhancement compared to the other samples. These results are consistent with studies finding AGN activity enhancement at this merger stage \citep{2013MNRAS.435.3627E,2023MNRAS.519.6149B, 2023ApJ...944..168L,2024MNRAS.528.5864B}. 

The differing average AGN activity enhancements for \textbf{a)} the entire merger sample without visual purification, \textbf{b)} the visually purified pair sample, \textbf{c)} the spectroscopically confirmed close pairs, and \textbf{d)} the visually purified postmerger sample shows both the ability and limitations of using a binary CNN for merger classification. Using a diverse sample of mergers to train a CNN will lead to mergers with diverse physical properties, appearances, and at various stages to be given high merger probabilities. Such a diverse sample will likely lead to statistically mixed physical results. Additional visual purification on CNN-based samples will lead to more clear results, which has been seen in previous studies \citep{2021MNRAS.504..372B}.

\subsection{Spectroscopically confirmed pairs}
\begin{figure}[h!]
    \centering
    \includegraphics[width=0.5\textwidth]{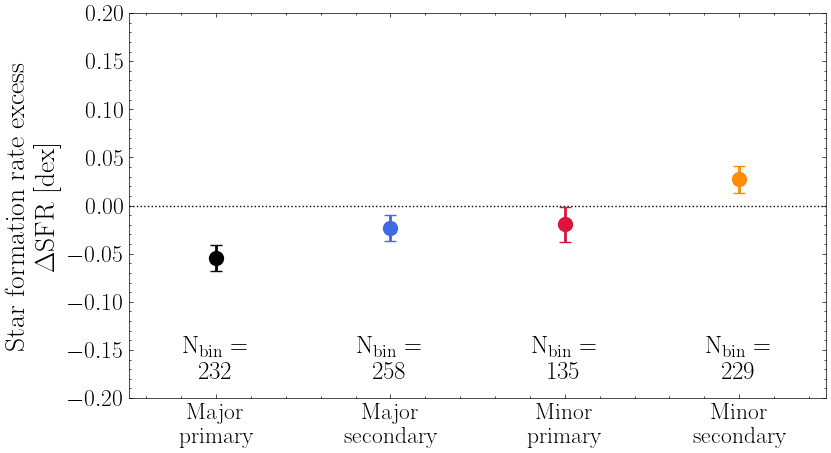}
    \caption{The same as Fig. \ref{fig:sfrenhancement}, but for spectroscopically confirmed major merger primary and secondary galaxies, and minor merger primary and secondary galaxies. We find that none of the samples have a pronounced enhancement in SFR.}
    \label{fig:sfrmajorminor}
\end{figure}

\begin{figure}[h!]
    \centering
    \includegraphics[width=0.5\textwidth]{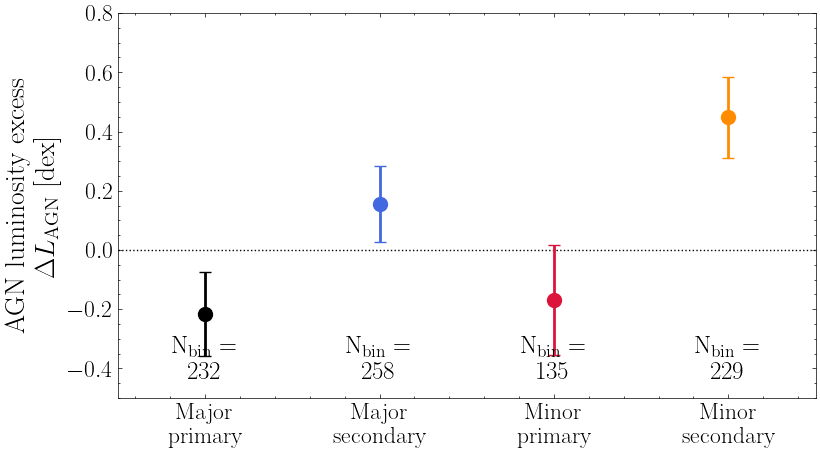}
    \caption{The same as Fig. \ref{fig:sfrmajorminor}, but for $L_{\mathrm{AGN}}$ enhancement. Similar to $f_{\mathrm{AGN}}$, we find that $L_{\mathrm{AGN}}$ is enhanced in the secondary galaxy and suppressed in the primary galaxy for both major and minor merger pairs.}
    \label{fig:lagnmajorminor}
\end{figure}

To compare with previous studies finding enhancements among pairs, we have conducted an additional investigation of SFR, $f_{\mathrm{AGN}}$ and $L_{\mathrm{AGN}}$ enhancement among spectroscopically confirmed pairs. We follow \citet{2013MNRAS.431..167R,2015MNRAS.452..616D} and consider pairs to be galaxies within $20 \mathrm{kpc} h^{-1}$ of each other and $\Delta v < 500 \mathrm{km/s}$. Of these pairs, we distinguished between major merger and minor merger pairs, with pairs with mass ratio $<$ 1:3 a major merger and pairs with mass ratio between 1:3 and 1:10 to be minor mergers, and the galaxy with greater $M_*$ of the pair considered the primary and lower $M_*$ the secondary. Figures \ref{fig:sfrmajorminor} to shows \ref{fig:lagnmajorminor} the results of these investigations.
We find that the primary galaxies for major mergers show a slight suppression, with $\Delta\mathrm{SFR}=-0.055\pm0.013$ dex. The secondary galaxies for major mergers also shows a slight suppression, with $\Delta\mathrm{SFR}=-0.023\pm0.013$ dex. For minor mergers, the primary galaxy shows a suppression of $\Delta\mathrm{SFR}=-0.020\pm0.018$ dex, and the secondary galaxy  is slightly enhanced, with $\Delta\mathrm{SFR}=0.027\pm0.014$ dex. We find that none of the samples show a pronounced enhancement or suppression in SFR, consistent with the results using machine-learning identified pair candidates, but not in line with the results of \citet{2015MNRAS.452..616D}. We attribute these results to the time-averaged SFRs obtained by \textsc{ProSpect}, in addition to possible effects on the SFR caused by the AGN.

\begin{figure}[h!]
    \centering
    \includegraphics[width=0.5\textwidth]{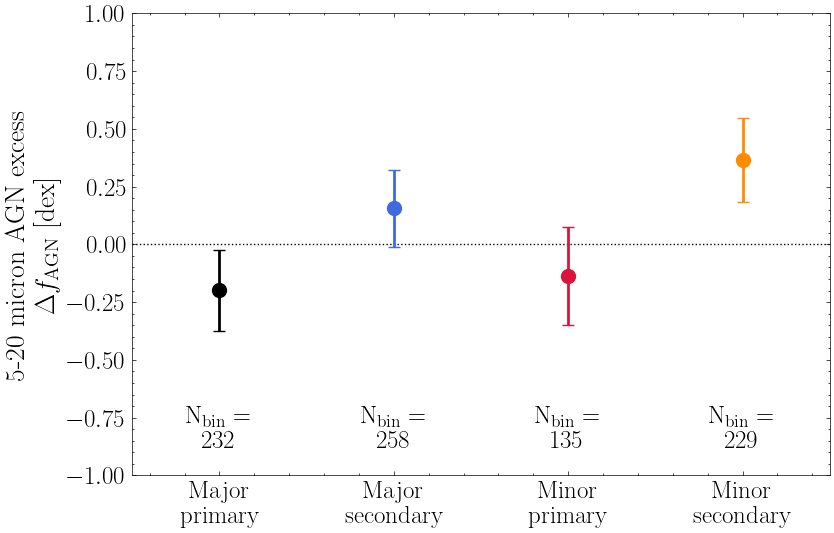}
    \caption{The same as Fig. \ref{fig:sfrmajorminor}, but for $f_{\mathrm{AGN}}$ enhancement. We find that $f_{\mathrm{AGN}}$ is enhanced in the secondary galaxy and suppressed in the primary galaxies for both major and minor merger pairs.}
    \label{fig:fagnmajorminor}
\end{figure}

For AGN enhancements, Fig. \ref{fig:fagnmajorminor} shows the $f_{\mathrm{AGN}}$ enhancement and Fig. \ref{fig:lagnmajorminor} shows the $L_{\mathrm{AGN}}$ enhancement. For $f_{\mathrm{AGN}}$ enhancement, the major mergers show a suppression of $\Delta\mathrm{SFR}=-0.199\pm0.177$ dex for the primary galaxies, and an enhancement of $\Delta\mathrm{SFR}=0.157\pm0.167$ dex for the secondary galaxies. For minor mergers, the primary galaxy shows a suppression of $\Delta\mathrm{SFR}=-0.136\pm0.211$ dex, and the secondary galaxy  is enhanced, with $\Delta\mathrm{SFR}=0.366\pm0.181$ dex. We find similar qualitative behaviour in $L_{\mathrm{AGN}}$, with a suppression in the primary of $\Delta\mathrm{SFR}=-0.216\pm0.143$ dex and an enhancement of $\Delta\mathrm{SFR}=0.155\pm0.128$ dex in the secondary for major mergers. Similarly for minor mergers, the primary galaxy has a suppression of $\Delta\mathrm{SFR}=-0.168\pm0.185$ dex, and the secondary galaxy has an enhancement of $\Delta\mathrm{SFR}=0.447\pm0.138$ dex.

In both mass ratio samples, we find that the secondary galaxy has an increase in AGN activity while the primary galaxy has a decrease in AGN activity compared to their controls. The enhancements in the secondary galaxy, greater in the minor merger pairs, are consistent with \citet{2015MNRAS.447.2123C}, which finds that black hole growth and AGN activity are more strongly affected by merger activity in the secondary galaxy. With regards to the primary galaxies, while we see a suppression in both $f_\mathrm{AGN}$ and $L_\mathrm{AGN}$, it does not necessarily equate to a decrease in AGN instance among merger pairs. An alternative mechanism for AGN triggering may exist, resulting in the control sample having enhanced AGN properties. This aligns with \citet{2012ApJ...751...72D, 2019MNRAS.489.4016S}, which finds that at $z > 1$, the dominant mechanism for AGN triggering is secular mechanisms, even though merger-triggered AGNs still exist.

\section{Conclusions}
\label{section:Conclusions}

In this work, we investigate the relationship between merger activity, star formation and AGN activity by investigating star formation enhancement and AGN activity enhancement between CNN identified mergers and controls from the HSC-SSP cross-matched with GAMA. We use galaxy and AGN properties, specifically stellar mass, star formation rate, AGN flux contribution ($f_{\mathrm{AGN}}$), and AGN luminosity ($L_{\mathrm{AGN}}$) which were measured by the SED fitting code \textsc{ProSpect}, a unique code which simultaneously obtains AGN and stellar properties. To identify mergers, we fine-tune the pre-trained publicly available deep representation learning model Zoobot using images and labels based on merger probabilities in the Galaxy Cruise project, and use the fine-tuned model to make merger predictions on HSC-SSP \textit{gri}-band images.

Our main findings include the following:
   \begin{enumerate}
      \item The merger probability distribution of our galaxies finds that over half of the galaxies are given low merger probabilities ($<0.3$), and about 10 \% of galaxies are given high merger probabilities ($>0.8$). 
      \item Visual inspection of 3000 galaxies finds 259 pairs and 100 postmergers. These results highlight both the strengths and weaknesses of binary classification. The model is able to identify mergers with diverse morphologies, indicating the ability to find mergers at various stages and with various physical properties. However, such a diverse sample may lead to statistically mixed results which may not draw accurate conclusions about the merger process.
      \item When investigating whether merger candidates (merger probability $>0.8$) have an enhancement in SFR, ratio of AGN flux to host flux ($f_{\mathrm{AGN}}$), and AGN luminosity ($L_{\mathrm{AGN}}$) relative to their controls (merger probability $<0.3$), we find very little enhancement in SF or AGN activity, with $\Delta\mathrm{SFR}=-0.009\pm0.003$ dex, $f_{\mathrm{AGN}}=-0.010\pm0.033$ dex, and $\Delta L_{\mathrm{AGN}}=0.002\pm0.025$ dex.
      These are not to suggest that mergers do not enhance star formation, but instead due to the diverse sample of mergers, as well as the SFRs being time-averaged over 100 Myr and secular star formation mechanisms. Similarly for the lack of enhancement in AGN properties, this is likely due to the diverse sample of mergers and possible alternative mechanisms for AGN triggering.
      \item When conducting the same comparison on the visually confirmed pairs and postmergers, we find that while the star formation rate enhancement remains mild, the enhancement in AGN activity is more pronounced in the visual purified samples, in particular the postmerger sample showing an AGN luminosity enhancement of $\Delta L_{\mathrm{AGN}} = 0.329\pm0.195$ dex, likely due to enhanced SMBH accretion rates at the postmerger stage.
\end{enumerate}

Our results present a cautionary tale when investigating the merger-SFR-AGN connection, particularly when using longer timescale SFR and AGN tracers. Further decomposition of the AGN and merger samples will be required to understand if there is a dependence on other physical properties on the merger-AGN relation. Future works will investigate dependencies on properties such as stellar mass, AGN luminosity, and environment.

\begin{acknowledgements}
We would first and foremost like to thank the native Hawaiians for sharing the Maunakea mountain, a place of cultural, historical, and natural significance, allowing us access to a beautiful view of the Universe.
Next, we would like to thank the anonymous referee for their comments and feedback in greatly improving the quality of this work.
KO was supported by JSPS KAKENHI Grant Number JP23KJ1089. CB gratefully acknowledges support from the Forrest Research Foundation. The Hyper Suprime-Cam (HSC) collaboration includes the astronomical communities of Japan and Taiwan, and Princeton University.  The HSC instrumentation and software were developed by the National Astronomical Observatory of Japan (NAOJ), the Kavli Institute for the Physics and Mathematics of the Universe (Kavli IPMU), the University of Tokyo, the High Energy Accelerator Research Organization (KEK), the Academia Sinica Institute for Astronomy and Astrophysics in Taiwan (ASIAA), and Princeton University. Funding was contributed by the FIRST program from the Japanese Cabinet Office, the Ministry of Education, Culture, Sports, Science and Technology (MEXT), the Japan Society for the Promotion of Science (JSPS), Japan Science and Technology Agency  (JST), the Toray Science  Foundation, NAOJ, Kavli IPMU, KEK, ASIAA, and Princeton University.

This paper is based [in part] on data collected at the Subaru Telescope and retrieved from the HSC data archive system, which is operated by Subaru Telescope and Astronomy Data Center (ADC) at NAOJ. Data analysis was in part carried out with the cooperation of Center for Computational Astrophysics (CfCA) at NAOJ. We are honored and grateful for the opportunity of observing the Universe from Maunakea, which has the cultural, historical and natural significance in Hawaii.

This paper makes use of software developed for Vera C. Rubin Observatory. We thank the Rubin Observatory for making their code available as free software at http://pipelines.lsst.io/.

\end{acknowledgements}

%
%

\bibliography{References}
\bibliographystyle{aasjournal}

\appendix
\section{Merger Mosaics}
\label{appendix: A}

In this section, we show the 20 image mosaics of mergers (merger probability $>0.8$), controls (merger probability $<0.3$), visually purified pairs, and visually purified postmergers, mentioned in Section \ref{Section:Samples}.

\figsetstart
\figsetnum{11}
\figsettitle{Mosaics}

\figsetgrpstart
\figsetgrpnum{8.1}
\figsetgrptitle{Merger mosaic}
\figsetplot{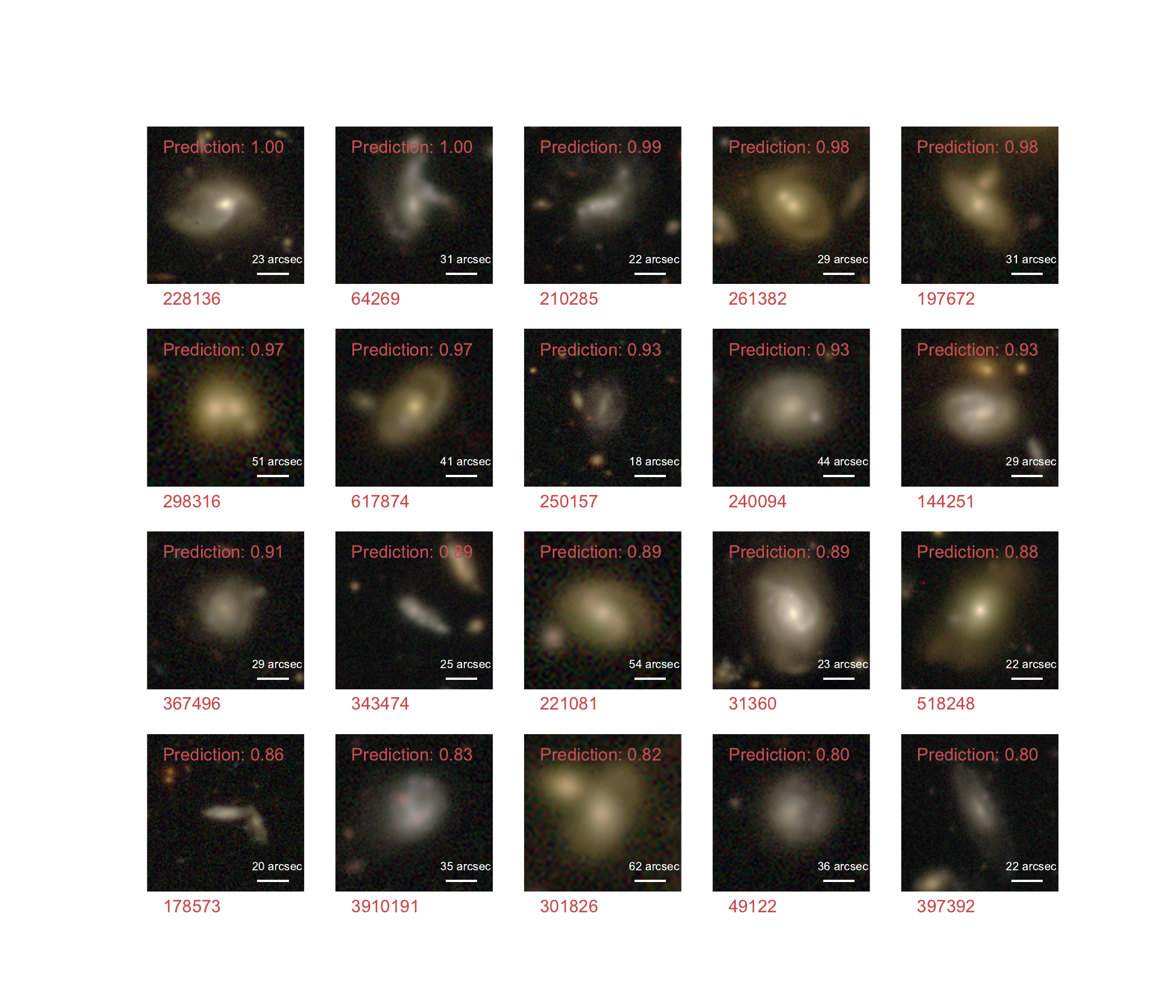}
\figsetgrpnote{20 randomly drawn examples of GAMA galaxies with a merger probability $>0.8$, with merger probabilities in descending order. The merger probabilities are indicated in the image and the GAMA ID below the image. The angular scales for the images are also available in the images in white text.}
\figsetgrpend

\begin{figure*}[!ht]
    \centering
    \hspace*{-3.5cm}\includegraphics[trim=0cm 3cm 0cm 0cm,scale=0.65]{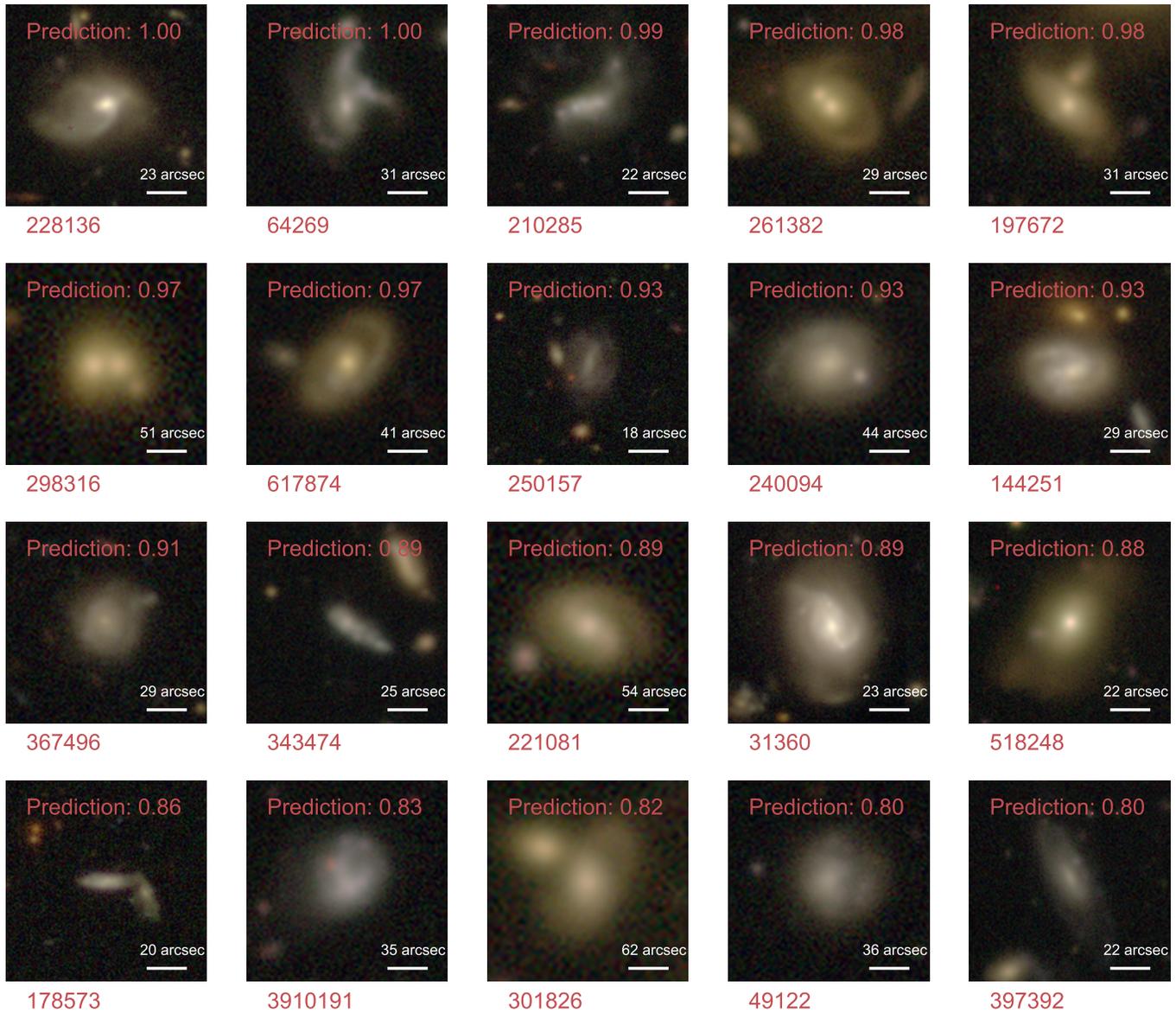}
    \caption{20 randomly drawn examples of GAMA galaxies with a merger probability $>0.8$, with merger probabilities in descending order. The merger probabilities are indicated in the image and the GAMA ID below the image. The angular scales for the images are also available in the images in white text. The complete figure set (4 images) is available in the online journal.}
    \label{fig:GAMAmergers}
\end{figure*}

\figsetgrpstart
\figsetgrpnum{8.2}
\figsetgrptitle{Non-merger mosaic}
\figsetplot{nonmergermosaic.pdf}
\figsetgrpnote{Same as Fig. \ref{fig:GAMAmergers} but for merger probability $<0.3$.}
\figsetgrpend

\figsetgrpstart
\figsetgrpnum{8.3}
\figsetgrptitle{Pair mosaic}
\figsetplot{pairmosaic0717.pdf}
\figsetgrpnote{Same as Fig. \ref{fig:GAMAmergers} but for the visually purified pair sample.}
\figsetgrpend

\figsetgrpstart
\figsetgrpnum{8.4}
\figsetgrptitle{postmerger mosaic}
\figsetplot{postmergermosaic0717.pdf}
\figsetgrpnote{Same as Fig. \ref{fig:GAMAmergers} but for the visually purified postmerger sample.}
\figsetgrpend

\figsetend

\newpage
\clearpage
\section{Enhancements with different controls}
\label{appendix: B}

In this section, we show Figs. \ref{fig:sfrenhancement} through \ref{fig:Lagnenhancement} where we sampled five, 20, and 100 controls for each merger.

\figsetstart
\figsetnum{12}
\figsettitle{Enhancements with different controls}

\figsetgrpstart
\figsetgrpnum{12.1}
\figsetgrptitle{SFR Enhancements 5 Controls}
\figsetplot{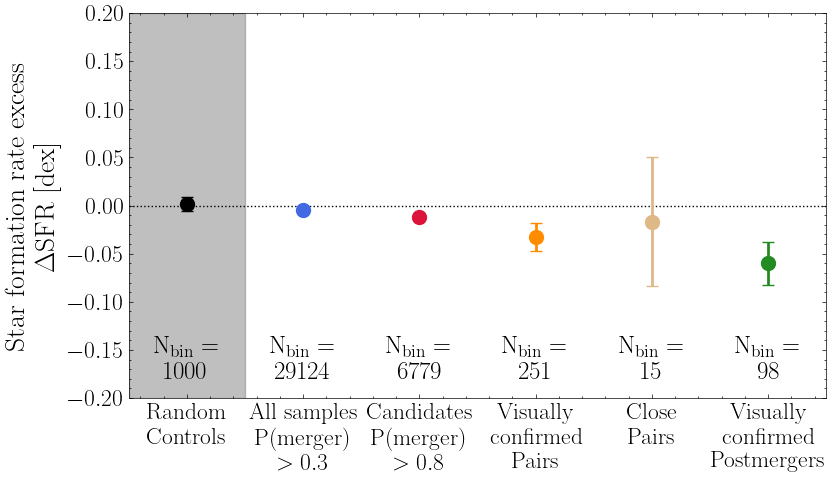}
\figsetgrpnote{Average star formation enhancements for each merger sample, with 5 controls. Enhancements for \textbf{a)} controls: $\Delta \mathrm{SFR}=0.002\pm0.007$ dex \textbf{b)} merger probability $>$0.3 sample: $\Delta \mathrm{SFR}=-0.004\pm0.013$ dex \textbf{c)} all merger candidates: $\Delta \mathrm{SFR}=-0.012\pm0.003$ dex \textbf{d)} pairs: $\Delta \mathrm{SFR}=-0.032\pm0.015$ dex \textbf{f)} close pairs: $\Delta \mathrm{SFR}=-0.017\pm0.067$ dex \textbf{f)} postmergers: $\Delta \mathrm{SFR}=-0.060\pm0.022$ dex.}
\figsetgrpend

\figsetend
\begin{figure}[ht]
    \centering
    \includegraphics[width=0.5\textwidth]{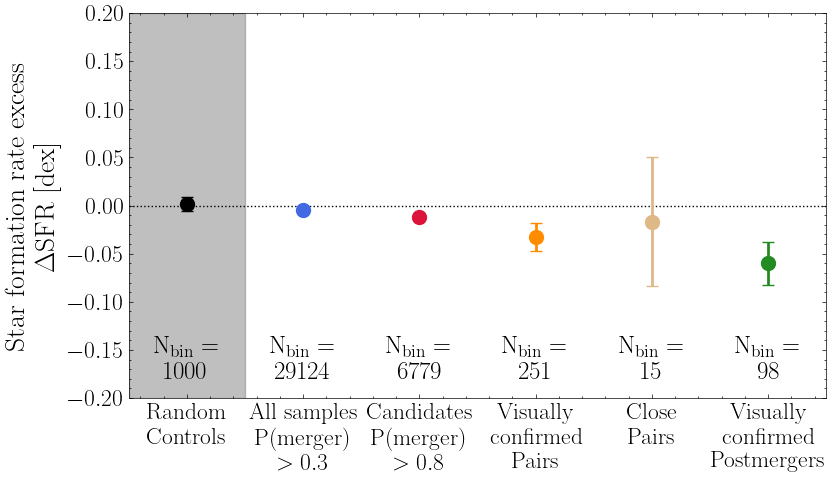}
    \caption{Average star formation enhancements for each merger sample, with 5 controls. Enhancements for \textbf{a)} controls: $\Delta \mathrm{SFR}=0.002\pm0.007$ dex \textbf{b)} merger probability $>$0.3 sample: $\Delta \mathrm{SFR}=-0.004\pm0.013$ dex \textbf{c)} all merger candidates: $\Delta \mathrm{SFR}=-0.012\pm0.003$ dex \textbf{d)} pairs: $\Delta \mathrm{SFR}=-0.032\pm0.015$ dex \textbf{f)} close pairs: $\Delta \mathrm{SFR}=-0.017\pm0.067$ dex \textbf{f)} postmergers: $\Delta \mathrm{SFR}=-0.060\pm0.022$ dex. The complete figure set (9 images) is available in the online journal.}
    \label{fig:sfrenhancement5}
\end{figure}

\end{document}